\documentclass[twocolumn]{revtex4-1}

\usepackage{graphicx}
\usepackage{graphics}
\usepackage{amssymb}
\usepackage{hyperref}
\usepackage{latexsym}
\usepackage{color}

\def\be{\begin{equation}}
\def\ee{\end{equation}}
\def\bea{\begin{eqnarray}}
\def\eea{\end{eqnarray}}

\begin{document}
\title{Glass-like slow dynamics in a colloidal solid with multiple ground states}
\author{Chandana Mondal$^{1}$, Smarajit Karmakar$^{2}$ and Surajit Sengupta$^{2}$}
\address{$^{1}$Centre for Advanced Materials, Indian Association for the Cultivation of Science, Jadavpur, Kolkata-700032, India.\\ $^{2}$TIFR Centre for Interdisciplinary Sciences, 21 Brundavan Colony, Narsingi, Hyderabad 500075, India.}

\begin{abstract}
We study the phase ordering dynamics of a two dimensional model colloidal solid using 
molecular dynamics simulations. The colloid particles interact with each other with a 
Hamaker potential modified by the presence of equatorial ``patches'' of attractive and 
negative regions. The total interaction potential between two such colloids is, 
therefore, strongly directional and has three-fold symmetry. Working in the canonical 
ensemble, we determine the tentative phase diagram in the density-temperature plane 
which features three distinct crystalline ground states viz, a low density honeycomb 
solid followed by a rectangular solid  at higher density, which eventually transforms 
to a close packed triangular structure as the density is increased further. We show 
that when cooled rapidly from the liquid phase along isochores, the system undergoes a 
transition to a ``strong glass'' while slow cooling gives rise to crystalline phases. 
We claim that geometrical frustration arising from the presence of many crystalline 
ground states causes glassy ordering and dynamics in this solid. Our results may be 
easily confirmed by suitable experiments on patchy colloids.
\end{abstract}
\maketitle

When a liquid is cooled sufficiently fast below its freezing temperature, molecules cannot adequately sample configuration space within the available time and the liquid fails to crystallise\cite{coolingrate1,coolingrate2} entering a supercooled state. On further cooling, many liquids eventually vitrify due to a large increase in viscosity and the associated structural relaxation time\cite{Debendetti-Stillinger,Ediger-Angell-Nagel}. In such a {\em glass forming liquid} the correlations between particle positions are observed to decay in two distinct steps: a short time $\beta$ decay resulting from the rattling motion of particles within cages formed by its neighbours, followed by $\alpha$ decay at longer times when the particles escape the cages\cite{2steprelaxation}. While there has been considerable work over decades\cite{Debendetti-Stillinger,liq-glass1,liq-glass2,liq-glass3,liq-glass4,liq-glass6,liq-glass7,liq-glass8,liq-glass9,liq-glass10,liq-glass11,liq-glass12}, several aspects of such glass transition phenomena are still not clearly understood. For example, many supercooled liquids, on the other hand,  crystallise\cite{Crystallization} when cooled further without going through a glass transition. What features of the molecular interactions and/or the quench protocol determine whether a liquid forms a glass or a crystal?

The origin of the glass transition and the associated slow dynamics has been the focus of rather intensive studies over the years. As a result, there exist several appealing explanations like caging induced memory effects\cite{caging1,caging2}, cooperative molecular motion\cite{cooperativity1,cooperativity2}, free volume\cite{freevolume1,freevolume2,freevolume3}, dynamical heterogeneity\cite{liq-glass6,haterogeneity}, and energetic and geometric frustration\cite{frustration1,frustration2,frustration3,frustration4,frustration5,frustration6,frustration7}. Among these, frustration is considered  to play the key role in glass transition. Hajime Tanaka and co-workers argued\cite{frustration2,frustration3,frustration7} that liquid-gas transition is controlled by the energetic frustration between long-range density ordering which favours normal liquid structures leading to crystallization and short range bond ordering which favours locally favoured structures due to complex many-body interactions. In Ref.\cite{Frank} Frank invoked a somewhat different role of frustration in glass transition. According to this theory, a liquid always prefer to freeze in its local structure which is different from the crystal structure, achievable through substantial molecular rearrangements. The local structure, being unable to tile the whole lattice, gives rise to geometrical frustration and eventually breaks the lattice into domains. Molecular rearrangements in different domains give rise to slow dynamics and triggers the glass transition.

In spite of these studies it is not yet obvious as to what enhances the glass forming ability of a material. For a monoatomic system with spherically symmetric interaction, it is very difficult to avoid crystallization even in computer simulations where higher (compared to experiments) cooling rate and finite system size disfavors it. However, if  geometrical frustration is introduced in terms of e.g., anisotropy in the the interaction or non-spherical shape\cite{Frenkel}, polydispersity\cite{polydisp1,polydisp2,polydisp4,polydisp5}, mixing of different types of particles\cite{mixing1,mixing2,mixing3} $\it{etc.}$, glass forming ability of the system is enhanced. In metallic alloys, multicomponent (more than three) systems with negative enthalphy of mixing and a large size ratio of components are known to be good glass formers\cite{metallicglass1,metallicglass2,metallicglass3,metallicglass4}. Sastry $\it{et \,al.}$ \cite{tetrahedrality} reported the condition under which a series of monoatomic glass formers were obtained tuning the tetrahedrality of a single component system $\it{viz.}$ the silicon potential. Interestingly, the best glass formers, at least within this model, were those at a value of tetrahedrality which supports more than one crystalline ground state.

Here we report on the glass forming ability of a single-component model colloid system in $2$d with anisotropic interactions which gives rise to many crystalline ground states which may coexist with each other for appropriate conditions. We show that, due to geometrical frustration arising from the competing crystalline states, the dynamics after a quench from the liquid is indistinguishable from slow glassy dynamics. 

The rest of this paper is organized as follows.
In Section \ref{model} we introduce the model of colloids which interact with patchy, angle-dependent interaction, properties common in network formers. We determine a tentative phase diagram of the model in Section \ref{patchy-pd} with the help of block-analysis technique\cite{CMSS,Block1,Block2} as well as the technique developed in\cite{MandH} for determining free energies of solid phases using simulations. In Section \ref{quench} we study the phase ordering and glass formation dynamics in our colloidal system. We conclude in Section \ref{summary} pointing out some directions of future work.\\

\section{The patchy colloid model}
\label{model}
\begin{figure}[h!]
\begin{center}
\includegraphics[width=2.6in]{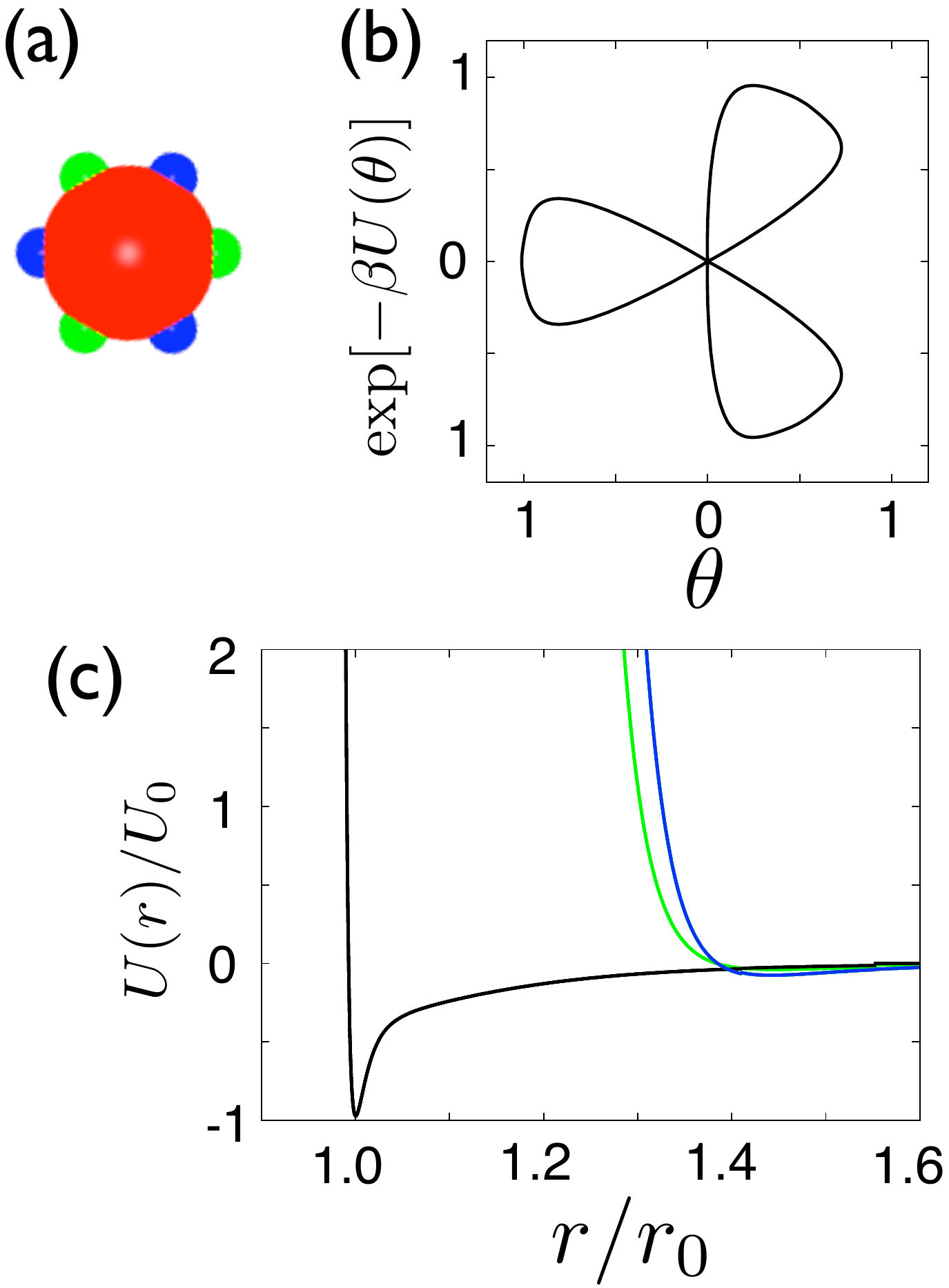}
\end{center}
\caption{(a) The patchy spherical colloidal particle (red circle) with patches (blue and green semi-circles) along the equator. (b) Boltzmann factor $exp(-\beta U(\theta))$, $\beta = 0.01$ as a function of the direction $\theta$ of ${\bf r_{ij}}$ at a fixed value of the magnitude $r$ of ${\bf r_{ij}}$ for two particles with relative phase angle $\omega_{ij} = \pi/3$. (c) Interaction energy, $U({\bf r_{ij}}, \omega_{ij})$, per molecule as a function of $r$ when two $type-2$ particles (black line), one $type-2$ and another $type-3$ particle (green line) and two $type-3$ particles (blue line) face each-other. Here $r_{0}=14.89575\sigma_{11}$ and $U_{0}=21A_{1,1}$.}
\label{Fig.14}
\end{figure}

We construct a model of colloidal particles with patchy, angle-dependent interactions of Hamaker\cite{Hamaker} and Lennard-Jones (LJ) type as implemented in the MD simulation package LAMMPS\cite{LAMMPS} which we use for all MD simulations reported in this paper. In our model, each molecule consists of one central, large, spherical particle with six small equidistant patches of alternating types on it's equator. In Fig.\ref{Fig.14}a, we draw a schematic for the patchy colloid particle. We refer to the central particle shown as a large red circle as a $Type-1$ particle and the blue and green semicircular  patches on it's equator as $Type-2$ and $Type-3$ particles respectively.

The interaction between two $Type-1$ particles $(U_{1,1})$ is a Hamaker interaction\cite{Hamaker}. The interaction between a $type-1$ particle and a $type-2/type-3$ particle is the interaction between a large size colloid particle and a solvent LJ particle. Two $type-3$ particles or a $type-2$ particle and a $type-3$ particle interact with simple LJ interaction. Details of the interactions are described in the appendix below. 
The sizes and interaction strengths between $type-1$, $type-2$ and $type-3$ particles are also tabulated in the appendix. All quantities in the table are expressed in reduced units. We choose the unit of length and energy are $\sigma_{11}$ and $A_{11}$. Also we choose the mass of each molecule $m=1$ without loss of generality.

We calculate the interaction energy between two patchy colloids $i$ and $j$,  $U({\bf r_{ij}},\omega_{ij})$, where ${\bf r_{ij}}={\bf r_{i}-r_{j}}$, $r=|\bf{r_{ij}}|$ is the distance and $\theta$ is the relative angle which the vector pointing from particle $i$ to $j$ makes with the $x$ axis. 
In addition to $\theta$, we define a relative phase angle between two patchy particles as $\omega_{ij}$ such that when $\omega_{ij}=\pi/3$, two $type-2$ particles face each other at $\theta=\pi/3$ and two $type-3$ particles face each other at $\theta=0$. Whereas, for $\omega_{ij}=0$, a $type-2$ particle and another $type-3$ particle face each other. A polar plot of the Boltzmann factor $exp[-\beta U(\theta)]$  as a function of the direction $\theta$ of ${\bf r_{ij}}$ at  a fixed value of the magnitude of ${\bf r_{ij}}$ and $\omega_{ij} = \pi/3$ is shown in Fig.\ref{Fig.14}b. Evidently, the angular dependence of the interaction energy shows a strong, in-plane, three-fold symmetry. The deepest minima is obtained at $r=r_0 (=1.89575\sigma_{11})$ when $type-2$ particles face each other. Plots of $U({\bf r_{ij}},\omega_{ij})$ for various $\omega_{ij}$ are shown in Fig.\ref{Fig.14}c. The interaction energy is calculated as a function of $r$, for three distinct cases {\it i.e.} when (A) $(\omega_{ij} = \pi/3, \theta = \pi/3)$, (B) $(\omega_{ij} = \pi/3, \theta = 0)$ and (C) $(\omega_{ij} = 0, \theta = 0)$. The interaction for all the three cases have a van der Waals form with attractive minima. Note that, the deepest minimum is obtained for case (A). In the next section we  determine the phase diagram in the density-temperature plane.

\section{Determination of the phase diagram}
\label{patchy-pd}

\begin{figure}[h!]
\begin{center}
\includegraphics[width=3.2in]{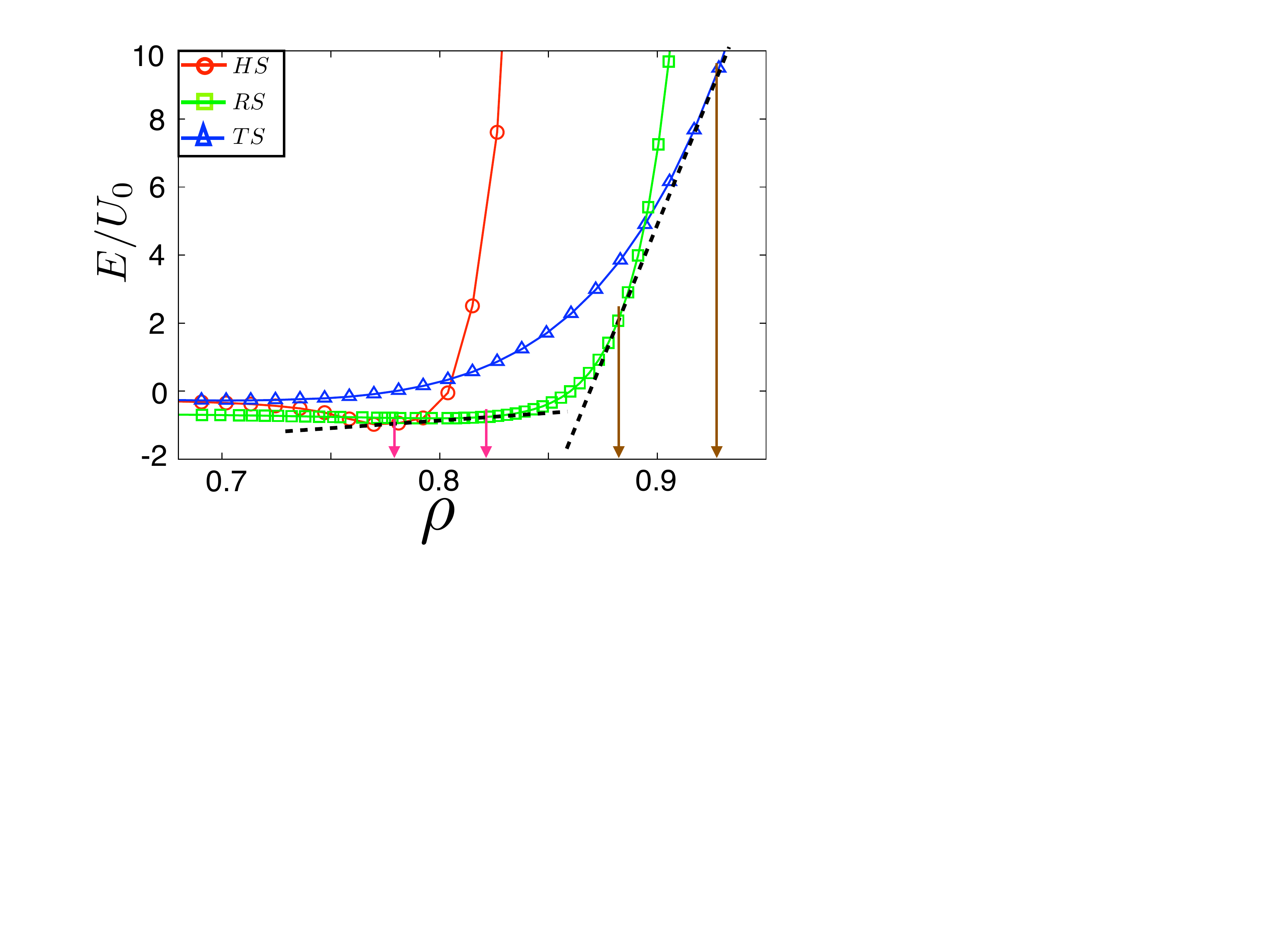}
\end{center}
\caption{Interaction energy per molecule minimized {\it w. r. t} the phase angle, $\omega_{ij}$ plotted as a function of density for honeycomb lattice (red circles and line), rectangular lattice (green rectangles and line), and triangular lattice (blue triangles and line). Density $\rho$ is scaled by $r_{0}^2$. The dashed lines are the common tangent to the associated energy-curves. The regions between the magenta arrows mark the HS+RS phase and the region between the mocha arrows mark the RS+TS phase.}
\label{Fig.24}
\end{figure}

In this section we study different phases of the system and determine the density-temperature phase diagram in two dimensions. Three-fold symmetry in the effective interaction (Fig.\ref{Fig.14}b) suggests that this model system should consist of a low-density open honeycomb solid (HS)\cite{CMSS} with a coordination number $3$, as ground state. At very high density, the system forms, as expected, a closed packed triangular solid (TS) state with coordination number $6$. At the intermediate densities the rectangular solid (RS) with a coordination number $4$, may form a ground state if energetically favourable. To investigate this we calculate the zero temperature energy minimised over the phase angle $\omega_{ij}$ as a function of density. Plot of the energy per molecule as a function $\rho$ is shown in Fig.\ref{Fig.24} for honeycomb lattice (red circles and line), rectangular lattice (green rectangles and line), and triangular lattice (blue triangles and line). 
The $T=0$ phase diagram is then obtained by drawing common tangents to  pairs of these curves.
%
%
%
%
%
In Fig.\ref{Fig.24} we see that the energy curve for RS has common tangent to the other two energy curves corresponding to HS and TS phase. This implies that RS is a ground state of the system and it may coexist with HS and TS phases as well. So, our model system consists of three crystalline ground states {\it viz.} HS, RS, and TS. The RS phase is stable at intermediate densities and it coexist with the HS phase at lower densities and with the TS phase at higher densities. The regions between the magenta arrows mark HS+RS phase coexistence and the region between the brown arrows mark the RS+TS phase. As is already known, existence of several high and low density ground states is a characteristic property of network-forming liquids \cite{polymorphism1,polymorphism2}.

For $T>0$, thermodynamic stability of different phases is determined by the free energy $F=E-TS$, where $S$ is the entropy of the system. At finite temperatures, the energy $E$ can be calculated from the ensemble average which is equal to the time average if the system is in equilibrium and is ergodic. But it is not trivial to evaluate $S$ directly as it cannot be expressed in terms of an ensemble average. 
In Ref.\cite{MandH} an accurate method is developed by Morris and Ho for determining approximate free energy of solid phases at temperature $T$ from a single simulation. Correlation functions available from the simulation are used to predict the upper bound on the entropy. This technique has been tested to give accurate result for a simple model of structural phase transition \cite{Jayee}.

We used this technique to estimate the free energy for our system at different density and temperature. We carry out constant ($N,A,T$) MD simulations where $N$ is the total number of molecules in the system, $A$ is the area(volume) of the system. We simulate the system at four different temperatures $T=0.1, 0.5, 1.0,$ and $1.5$ and at densities ranging from $\rho=0.68 $ to $\rho=1.0$. We simulate all the three kinds of solid taking $N=864$ for honeycomb lattice and rectangular lattice and $N=1024$ for triangular lattice. To achieve constant temperature, we use dissipative particle dynamics thermostat\cite{dpd-tstat1,dpd-tstat2}.
In each state we simulate the system for $2\times 10^8$ MD steps with an integration time-step $\delta t=5\times10^{-3}$.

\begin{figure}[h!]
\begin{center}
\vskip +0.2cm
\includegraphics[width=3.4in]{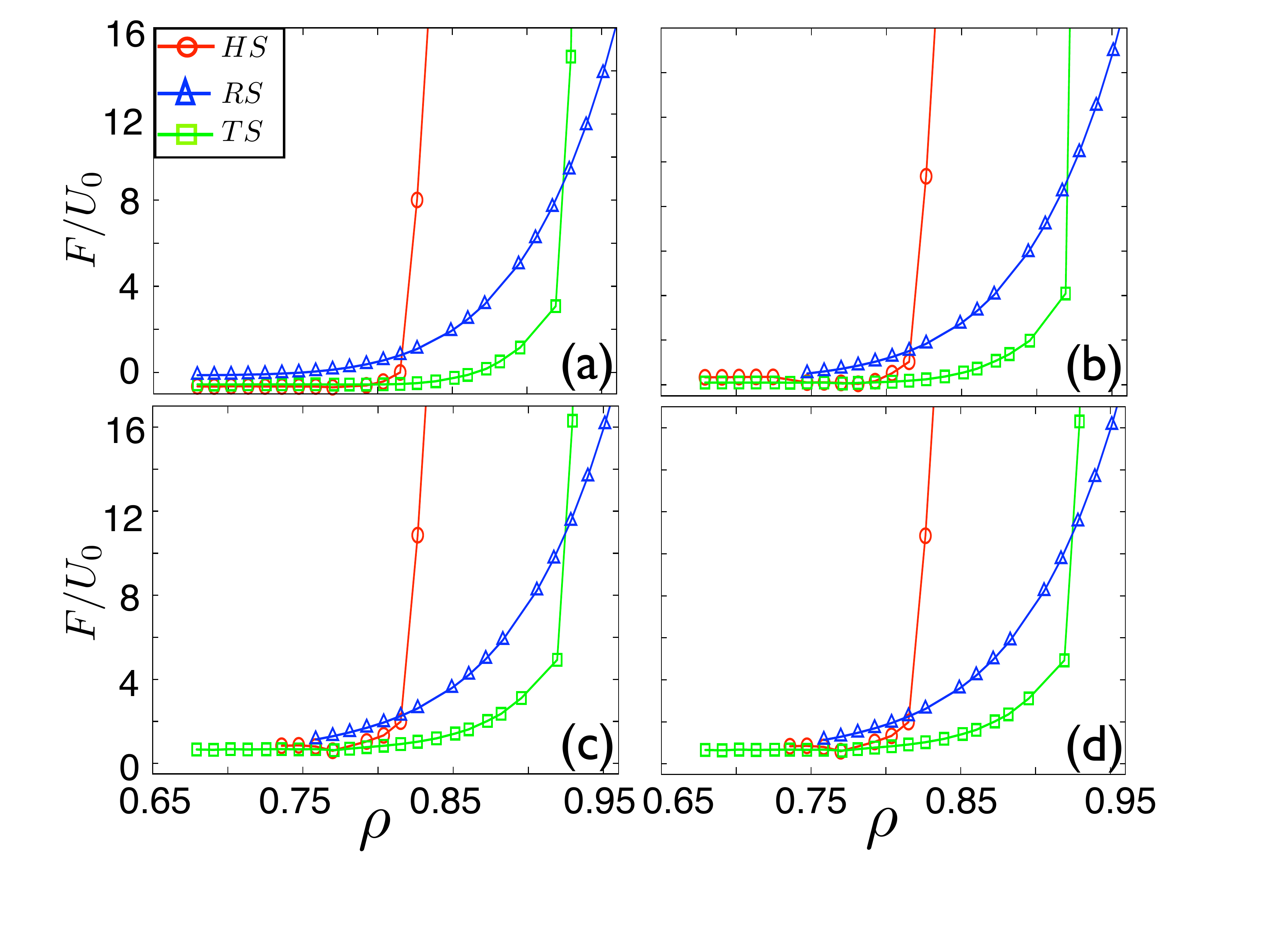}
\end{center}
\caption{Free energy per molecule as a function of density for honeycomb lattice (red circles), rectangular lattice (green rectangles), and triangular lattice (green squares) for temperature (a) $T=0.1$ and (b) $T=0.5$, (c) $T=1.0$ and (d) $T=1.5$.}
\label{Fig.34}
\end{figure}

In Fig.\ref{Fig.34} we plot the free energy curves for HS (red circles and line), RS (green squares and line), and TS (blue triangles and line) as a function of density for temperatures (a) $T=0.1$, (b) $T=0.5$, (c) $T=1.0$, and $T=1.5$. The coexistence regions between various solid phases are obtained by drawing common tangents (not shown for clarity of the picture) as is done for Fig.\ref{Fig.24}. 

\begin{figure}[h!]
\begin{center}
\includegraphics[width=2.6in]{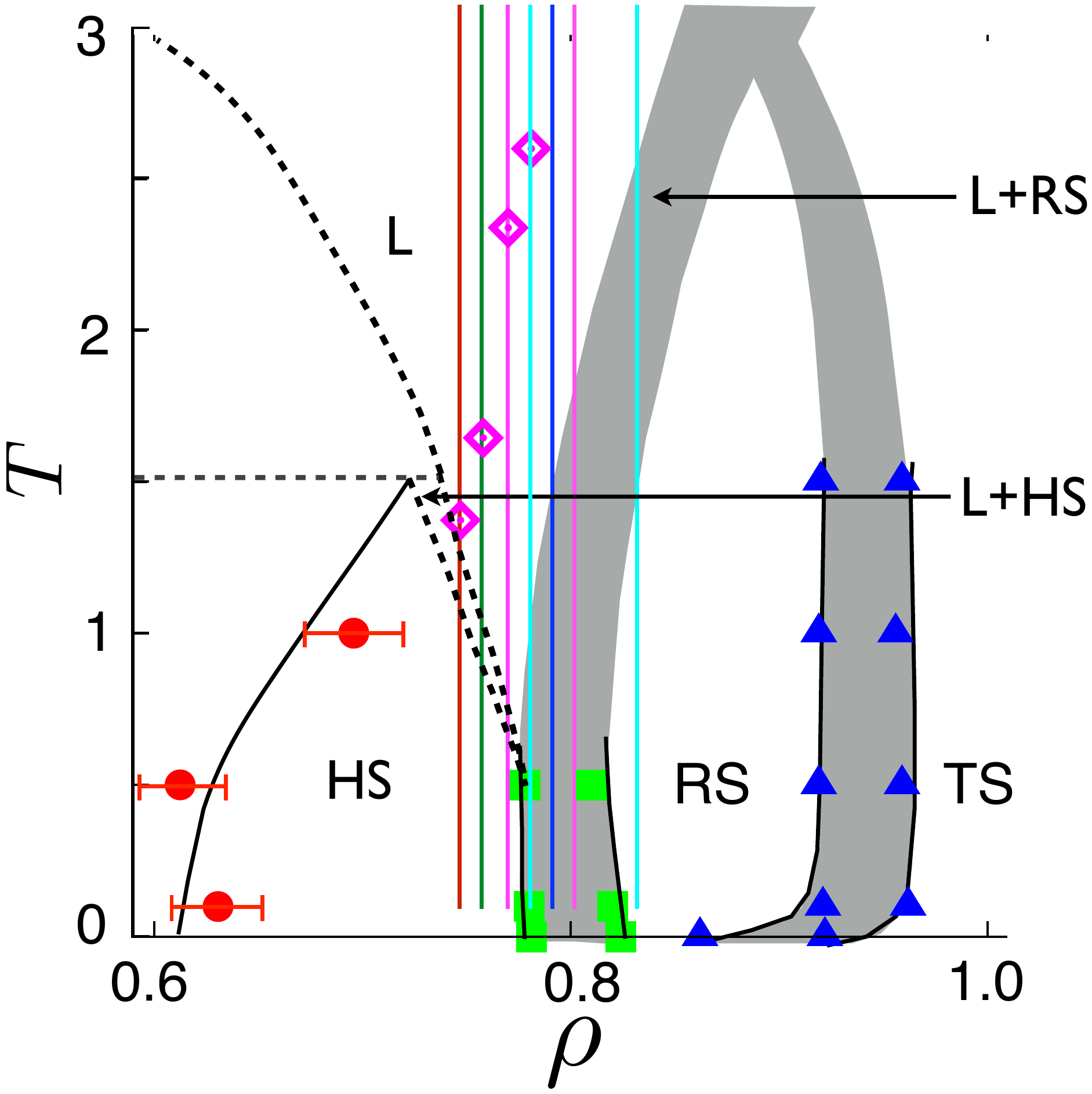}
\end{center}
\caption{Phase-diagram of the patchy-colloid model in $\rho-T$ plane obtained from MD simulations. The red circles plotted with error bar are obtained from block-analysis technique. The black line connecting these circles show the phase boundary of the honeycomb solid (HS) phase. Black dashed lines show expected phase boundaries of gas$+$liquid (G$+$L), liquid$+$rectangular solid (L$+$RS), and honeycomb solid$+$liquid phases. The green squares and the blue triangles, obtained from Morris and Ho free energy curves shown above, show the phase boundaries of the HS$+$RS phase and RS$+$TS phase respectively. The two phase coexistence regions are shaded in grey. The solid isochoric lines and the magenta rhombi will be explained later.}
\label{Fig.44}
\end{figure}

The points obtained by this method are shown in the phase diagram, Fig.\ref{Fig.44}. The green squares show the phase boundary for HS$+$RS coexistence phase while the blue triangles mark the phase boundary for RS$+$TS phase. At very high density TS phase is stable as it is the most closed packed structure in $2$d. To obtain the phase boundary for the HS phase {\it w.r.t.} the low density gas (G) phase, we use block analysis technique\cite{CMSS,Block1, Block2}. The red circles, obtained from block analysis, show the phase boundary for the HS phase {\it w.  r.  t.} the G phase. All the coexistence regions are shaded in grey. Phase boundaries of other coexistence phases like G$+$HS, G$+$liquid (L) and L+HS are not determined in this study though we have observed these phases in our simulations. The dashed lines are drawn to indicate the existence of these phases.  In this paper, we concentrate only on the phase boundaries for the three solid phases. Since the particles interact with patchy, angle-dependent interaction, and the model consists of several high and low density crystalline ground states like other network formers, the liquid state of our system is expected to show anomalous behaviour\cite{CMSS}. We do not study anomalies of the liquid phase in detail here.

\section{Quench dynamics}
\label{quench}
After determining the phase diagram of the patchy colloid system, we now analyse the dynamics of phase ordering after a quench (or slow cooling) from the liquid into the solid phases. It is well known that network forming liquids like silicon (Si) and silica (SiO$_2$) undergo a glass transition when cooled moderately rapidly to low temperatures. We study the glass-forming ability of our model colloid with patchy interactions and show that the presence of many competing crystalline ground states (see Fig.\ref{Fig.44}) frustrates perfect crystallization and leads to amorphous low temperature structures. For this reason we quench the system along different iso-chores which are shown in the phase-diagram: $\rho = 0.747, 0.758, 0.770, 0.781, 0.792, 0.804,$ and $0.838$ as red, green, magenta, cyan, blue, magenta, and cyan lines respectively. Details of simulations are given below.

\subsection{Simulation details}
We carry out molecular dynamics (MD) simulations\cite{LAMMPS}
with an integration time-step $\delta t = 5\times10^{-3}$. The model is simulated in constant $NAT$ ensemble and the system-temperature is kept fixed at the specified value using dissipative particles dynamics (DPD) thermostat\cite{dpd-tstat1,dpd-tstat2, dpd-LAMMPS}. First, we melt the system at a very high temperature $T=24.0$ to a fluid state and then anneal it step-by-step to $T=0.1$ along different iso-chores {\it viz.} $\rho = 0.747, 0.758, 0.770, 0.781, 0.792, 0.804,$ and $0.838$ which are indicated on the phase diagram, Fig.\ref{Fig.44}, as red, green, magenta, cyan, blue, magenta, and cyan lines respectively.

\subsection{Glassy dynamics}
 Though our model colloidal system admits three crystalline ground states {\it viz.} HS, RS, and TS separated by corresponding coexistence regions, homogeneous crystallization is prohibited and an amorphous, heterogeneous, polycrystalline or glass-like state is formed due to geometrical frustration and the high cooling rate in our MD simulations. The slowing down of dynamics of the system which is indicative of a glass transition, is exemplified by measuring a two-point correlation function {\it viz.} the intermediate scattering function which we describe below.

Self part of the intermediate scattering function is defined as,
\begin{eqnarray}
F_{k}(t,T) =  \frac{2}{N}\sum_{i=1}^{N/2}\exp(i\vec k . [\vec r_{i}(t) - \vec r_{i}(0)])   \nonumber \\
\label{scattf}
\end{eqnarray}
where, the average is over different time origins and the $|\vec{k}|$ 
corresponds to the $1^{st}$ peak position of the structure factor. 
The function $F_{k}(t,T)$ decays to zero at large times when system looses
its structural correlation with its initial configuration.

\begin{figure}[h!]
\begin{center}
\includegraphics[width=3.5 in]{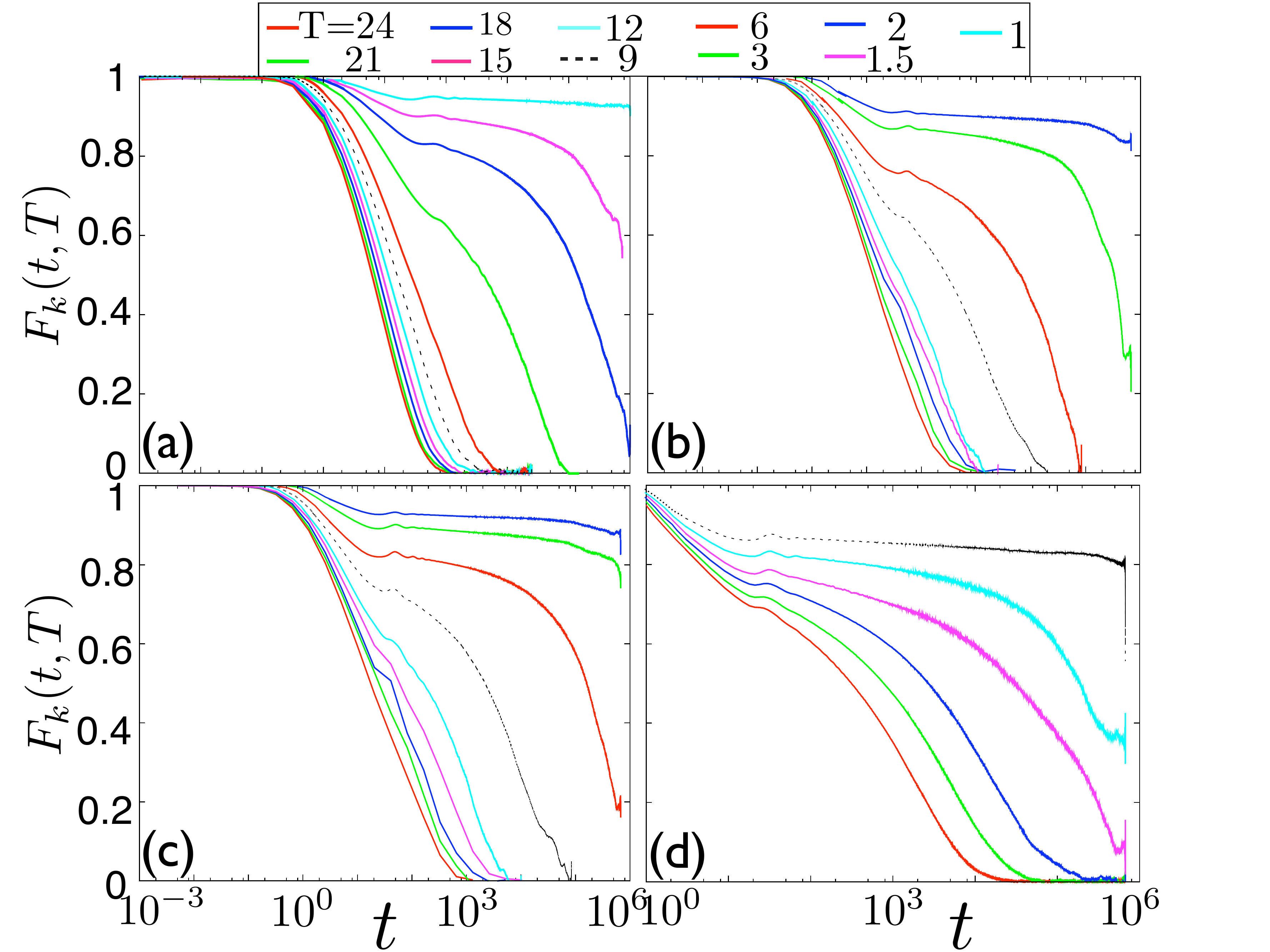}
\caption{Self part of the intermediate scattering function $F_{k}(t,T)$ as defined in Eqn.\ref{scattf} for $k(2\pi /a, 0)$, where $a$ is the lattice parameter, as a function of time $t$ for different temperatures $T$ at densities (a) $\rho=0.747$, (b) $\rho=0.792$, (c) $\rho=0.804$, (d) and $\rho=0.838$.} 
\label{Fig.26}
\end{center}
\end{figure}

We simulate the system quenching from the liquid into states with different densities corresponding to the crystalline ground states, and obtain $F_{k}(t,T)$ (see Fig.\ref{Fig.26}) as described in Eqn.\ref{scattf}. When  $F_k(t,T)$ decays to zero, we conclude that there are no persistent structural correlation present in the system. The relaxation time for {\it $\alpha$ relaxation}, $\tau_{\alpha}$, is estimated as the time at which $F_{k}(t,T)$ decays to $1/{\it e}$ of it's initial value at time $t=0$ {\it i.e.} $F_k(\tau_{\alpha}, T)=F_k(0,T)/e$. Below a certain temperature, which increases with increasing density, the relaxation time becomes so large that within the simulation time-scale (maximum run-length in our simulations is $2\times10^{8}$ MD steps with step-size $\delta t=5\times10^{-3}$) equilibrium cannot be reached.
This is a signature of glass transition with long-lived structural correlations which occurs at a temperature $T \leq T_g$. The slowing down of the system's dynamics is evident from Fig.\ref{Fig.26} where we plot $F_{k}(t,T)$ as a function of time $t$ obtained by quenching the system from $T=24.0$ to $T=0.1$ along four different iso-chores. At very high temperatures, the correlation function shows rapid exponential decay -- a characteristic of the liquid phase. As temperature is lowered a plateau develops demarcating the decay process into two regimes, a rapid decay at short times to the plateau where the relaxation is slow, followed by a long time decay after the plateau. The very slow decay regime during the plateau is called the {\it $\beta-regime$}.  A {\it boson peak} is also seen as shown in Fig.\ref{Fig.26}. The subsequent large time slow decay is called the {\it $\alpha$ relaxation regime}. All of these phenomena appear to indicate an approaching glass transition and the dynamics is indistinguishable from glassy dynamics. Note that the {\it $\beta-relaxation$} plateau becomes more prominent as $\rho$ is increased and parameters are chosen such that the system supports a mixture of HS and RS ground states.
 
\begin{figure}[h!]
\begin{center}
\includegraphics[width=3.4 in]{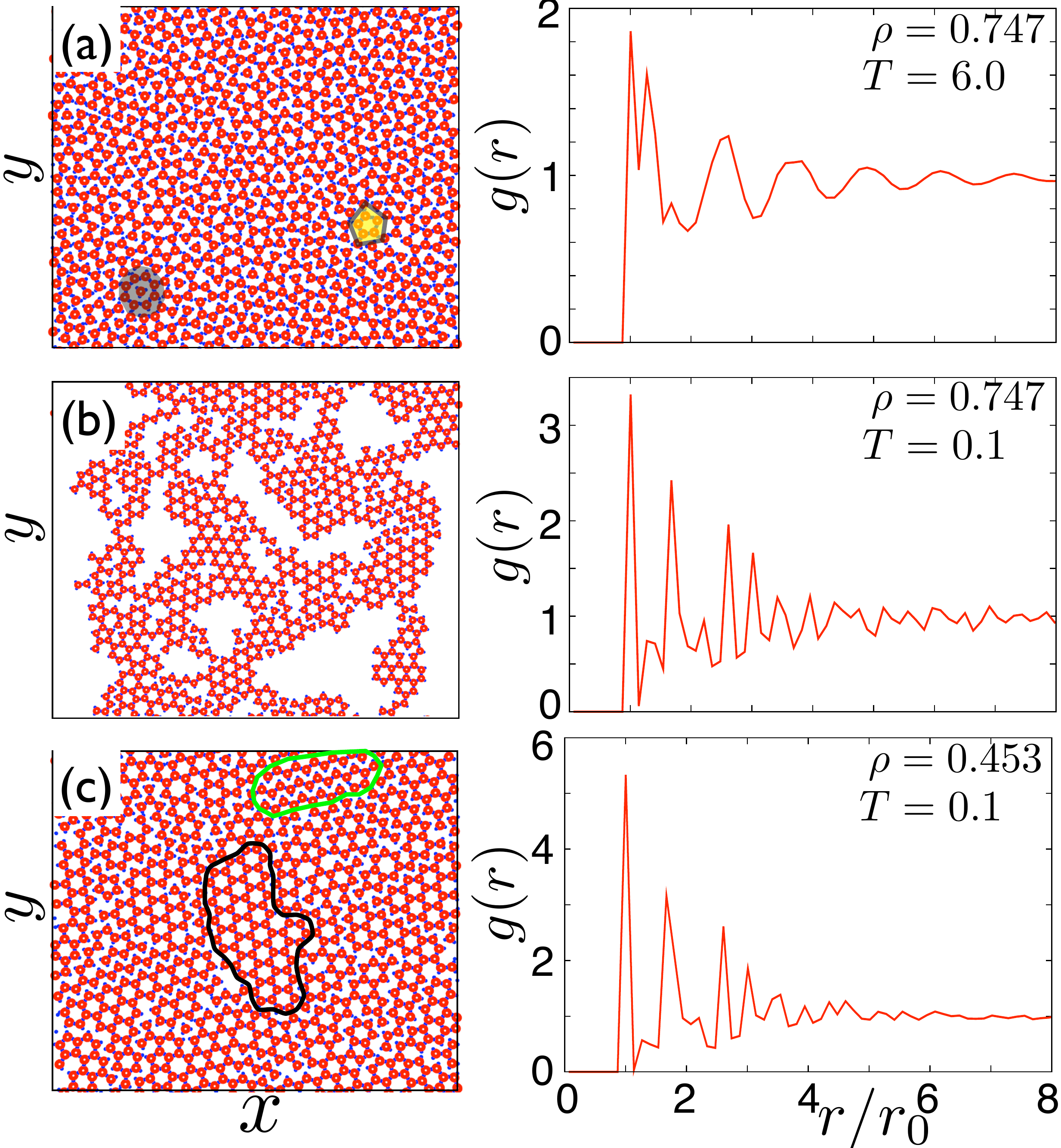}
\caption{Snapshot pictures of different phases of the system obtained from simulations for (a) $\rho=0.747, T=6.0$, (b) $\rho=0.747, T=0.1$, and (c) $\rho=0.453, T=0.1$. The yellow shadings highlights pentagons while the grey shadings show caging. The green boundary covers a crystallite region with four-fold symmetry and the black boundary covers a crystallite region with three-fold symmetry. Radial distribution function corresponding to the three phases are shown in the right.} 
\label{Fig.16}
\end{center}
\end{figure}

To give an idea about the microscopic structures of the different phases discussed here, we show in Fig.\ref{Fig.16}~(left) three snapshots: (a) $\rho = 0.747, T=6.0$, showing the liquid phase (b) $\rho=0.453, T=0.1$, showing G-HS coexistence, and (c) $\rho=0.747, T=0.1$, showing a typical glassy phase. In Fig.\ref{Fig.16}~(right) we show the corresponding radial distribution functions for these three phases. Note that, all the radial distribution functions show the $1^{st}$ peak at $r_{0}$, the minimum of the interaction potential. An interesting observation as is clearly seen from Fig.\ref{Fig.16}(a) is that the liquid phase contains many pentagonal and octagonal structures. Pentagonal structures, which are the locally favoured structures, frustrate crystallization as pentagons cannot tile $2$d space. One such pentagon is illustrated by yellow shading. The octagonal structures on the other hand (marked by grey shading), show regions where particles are trapped within cages formed by their neighbors. These particles can only move through cooperative movement of the cages which slows down the system's dynamics noticeably. The glassy state in Fig.\ref{Fig.16}(c) consists of small crystallites of three-fold symmetry (HS) (region within the black boundary) as well as four-fold symmetry (RS) (region within the green boundary) and other amorphous regions - a structure whose formation kinetics mimics that of a glass in our system.

\begin{figure}[h!]
\begin{center}
\includegraphics[width=3.0 in]{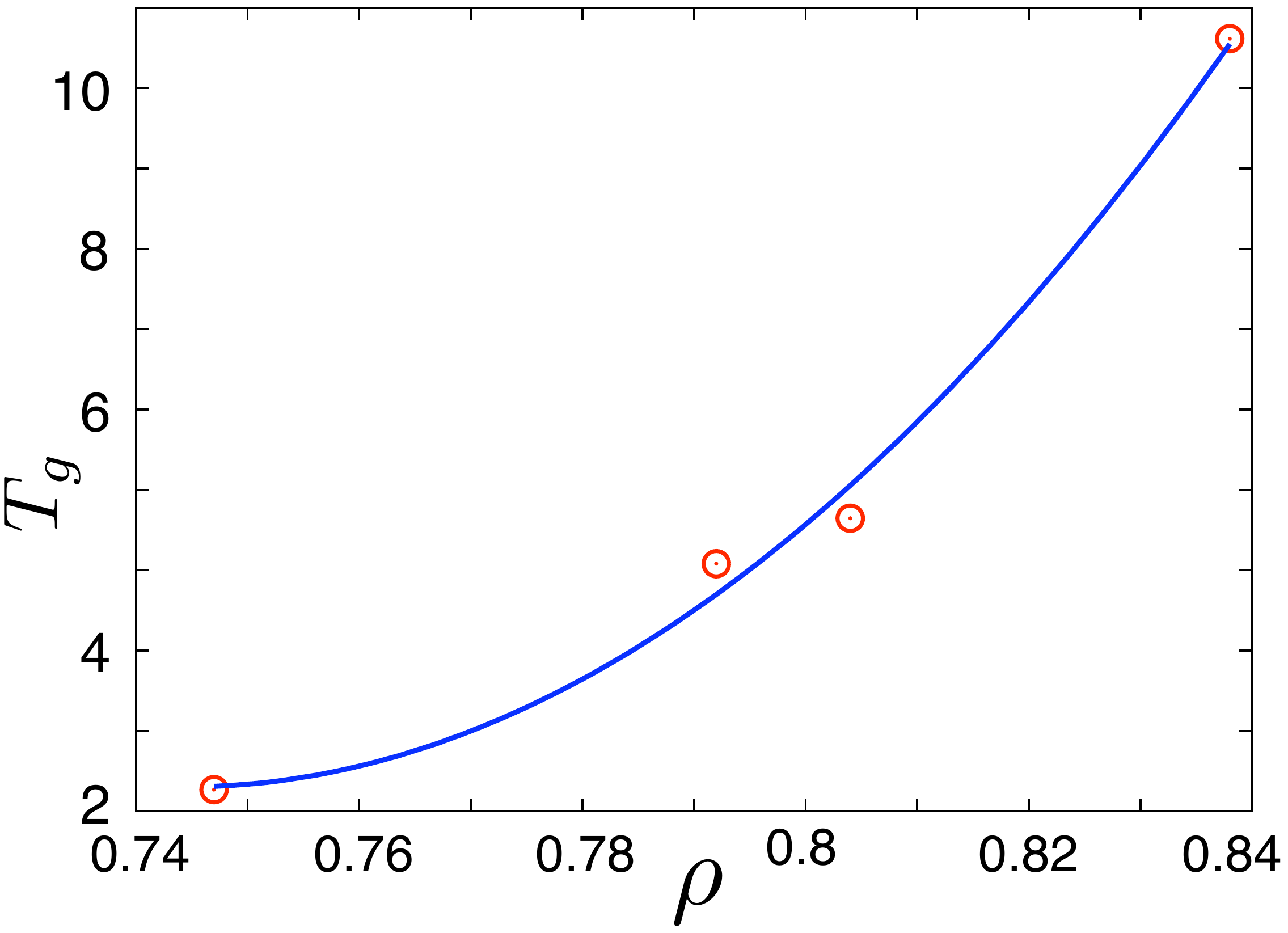}
\caption{The glass transition temperature $T_{g}$ as a function of density.} 
\label{Fig.36}
\end{center}
\end{figure}
 
\begin{figure}[h!]
\begin{center}
\includegraphics[width=3.0 in]{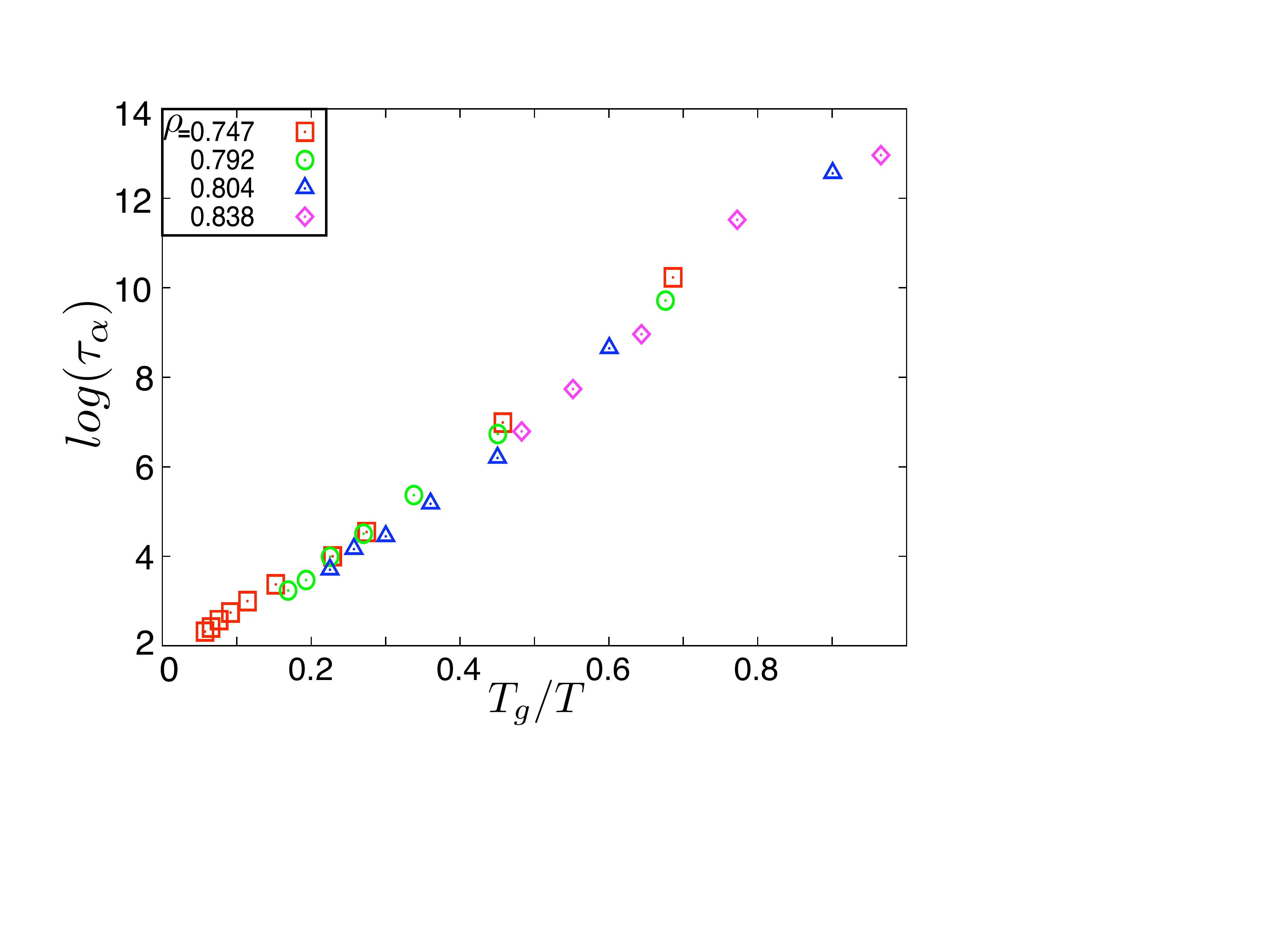}
\caption{Data collapse plot: plot of the log of the relaxation time $\tau_{\alpha}$ as a function of $T_g/T$ for $\rho=0.747$ as  red squares, $\rho=0.792$ as green circles, $\rho=0.804$ as blue triangles, and $\rho=0.838$ as pink rhombi. All the plots collapse to a single straight line.} 
\label{Fig.46}
\end{center}
\end{figure}

We measure the glass transition temperature $T_g$, roughly, as the temperature at which the relaxation time becomes $10^6$. $T_g$ is measured for all the densities and plotted in Fig.\ref{Fig.36} as red circles. The blue line, a polynomial fit to the data, shows a quadratic increase of $T_g$ with density. 
In Fig.\ref{Fig.46} we plot $\log(\tau_{\alpha})$ as a function of $T_g/T$ for four densities  $\rho = 0.747$ (red squares), $0.792$ (green circles), $0.804$ (blue triangles) and $0.838$ (pink rhombi). The data collapse nicely in Fig.\ref{Fig.46} in a single line. Linear nature of this plot implies that the temperature variation of the relaxation time is Arrhenius-like i.e. $\tau_{\alpha}\propto \exp(1/T)$ - the signature for a ``strong'' glass.

{\it Local structure:} The Arrhenius nature of the relaxation time implies that our system forms a strong glass\cite{strong} like other network forming liquids. The molecular interactions of our system has three-fold symmetry (see Fig.4.1b) which favours bond formation along these direction leading to strong orientational correlations in particle's position of the system. Though thermal fluctuations at very high temperatures, may reduce the three-fold coordinated structure, the system gains strong correlations gradually with the lowering of the temperature. The subtle changes in the structural organisation of the particles emerge from a study of the local structure of the system at different temperatures.

\begin{figure}[h!]
\begin{center}
\includegraphics[width=3.5 in]{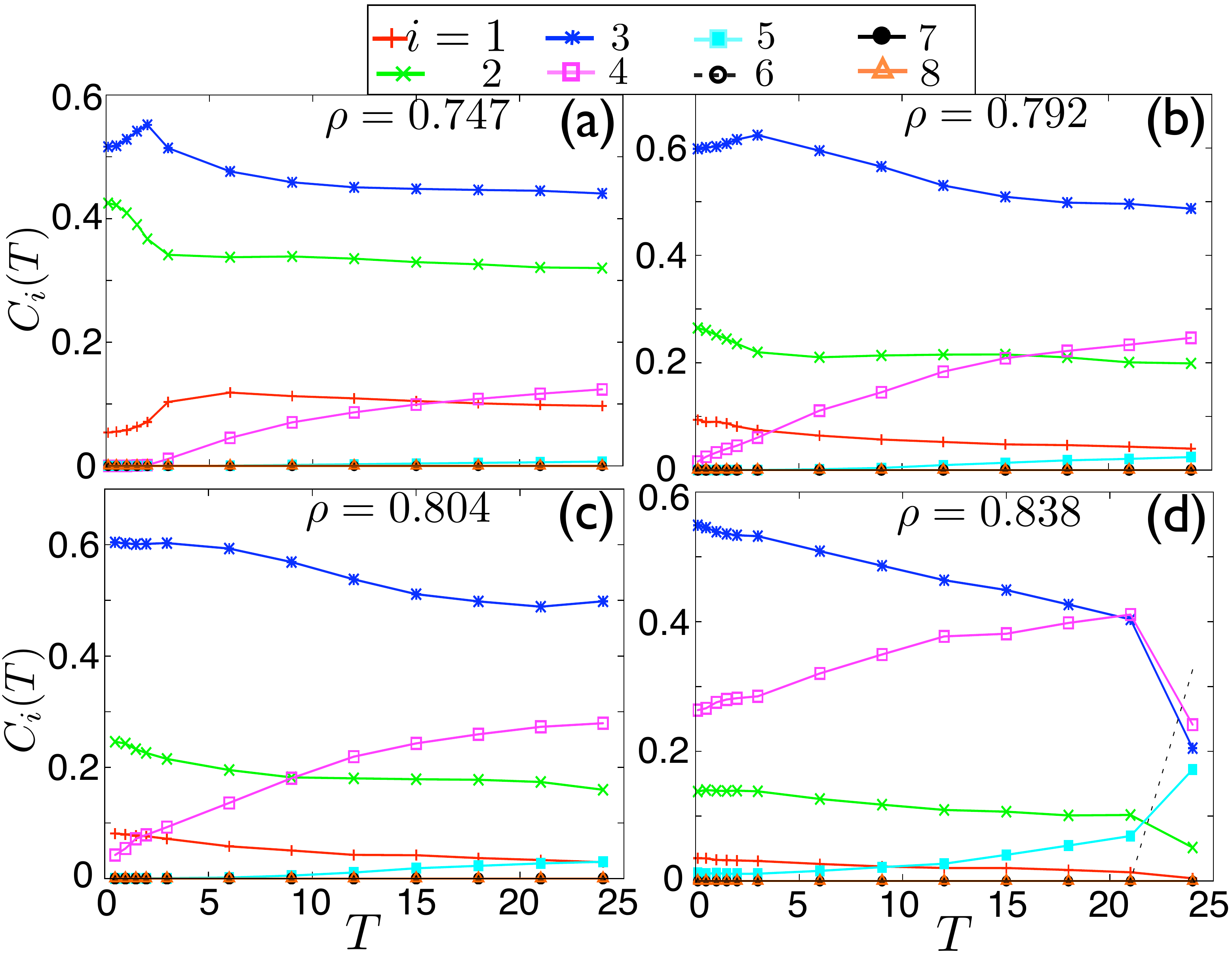}
\caption{Temperature dependence of the concentrations of quasi-species with $r_c/r_0=1.235$ for density (a) $\rho=0.747$, (b) $\rho=0.792$, (c) $\rho=0.804$, (d) and $\rho=0.838$.} 
\label{Fig.56}
\end{center}
\end{figure}

\begin{figure}[h!]
\begin{center}
\includegraphics[width=3.5 in]{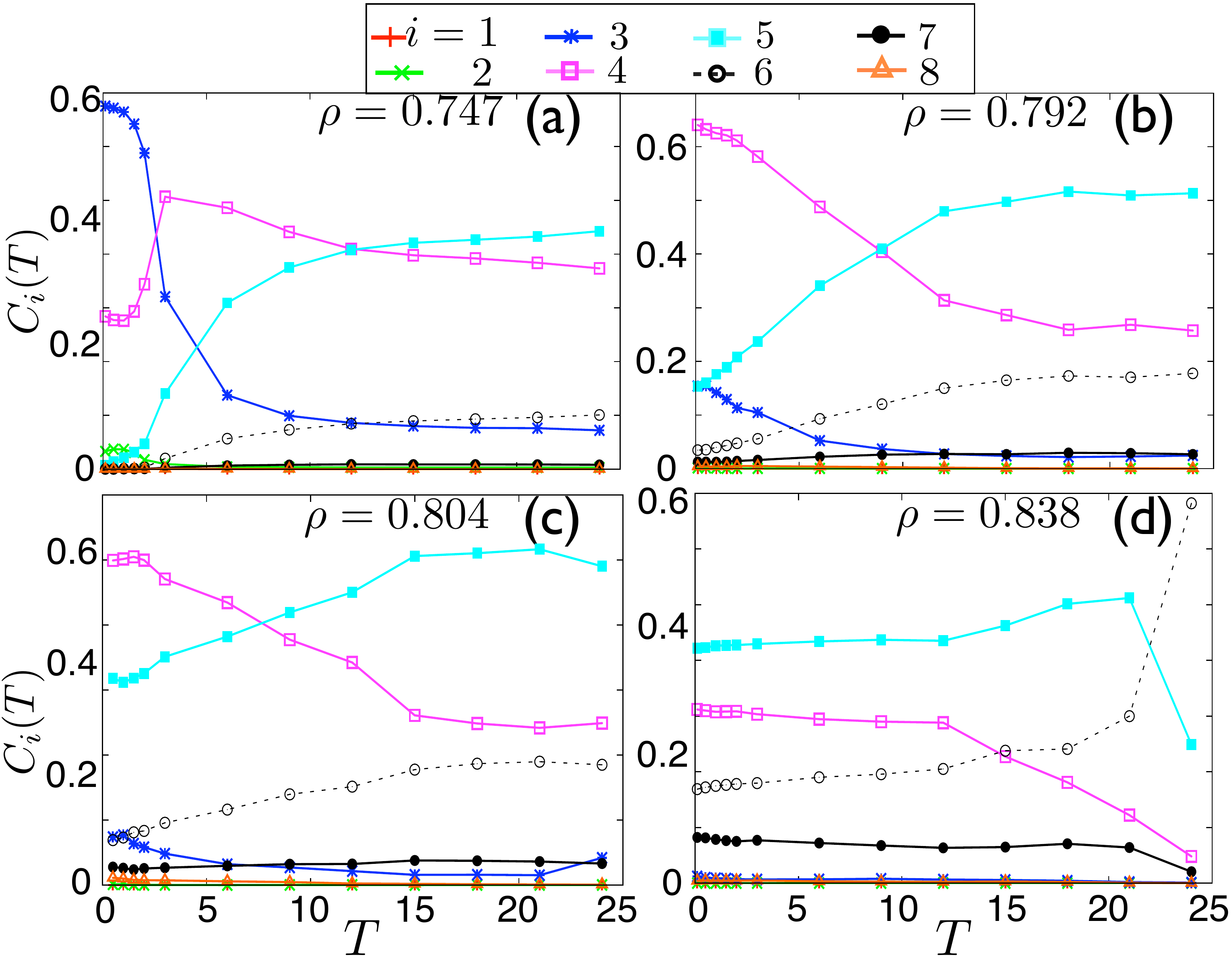}
\caption{Temperature dependence of the concentrations of quasi-species with $r_c/r_0=1.409$ for density (a) $\rho=0.747$, (b) $\rho=0.792$, (c) $\rho=0.804$, (d) and $\rho=0.838$.} 
\label{Fig.66}
\end{center}
\end{figure}
A very convenient and well used \cite{dos} measure of the local structure of a crystal is the calculation of it's density of species. Density of species which is generally denoted by $C_i(T)$ is the average (over time) density of those particles which has $i$ number of neighbors within a specified cutoff $r_c$ at temperature $T$. The value of $C_i(T)$ depends crucially on temperature $T$ as well as on the choice of $r_c$. If $r_c$ is set at the minima of the interaction potential ($r_c$ should be slightly greater than the potential minima to take into account fluctuation in particle positions at a finite temperature), then $C_i(T)$ gives the average density of particles with $i$ number of nearest neighbors. In case of a perfect HS with orientationally correlated three-fold symmetric particle positions, there is only one kind of species with $i=3$, while for RS $i=4$ and for TS $i=6$. $C_i(T)$. Polycrystalline or amorphous solids show distribution of $i$. We have calculated $C_i(T)$ for our system during temperature quenches. In Fig.\ref{Fig.56} and Fig.\ref{Fig.66} we show the temperature dependence of $C_i(T)$ for (a) $\rho=0.747$, (b) $\rho=0.792$, (c) $\rho=0.804$, and (d) $\rho=0.838$ for two different values of $r_{c}$ {\it viz.} $r_{c}=1.235r_0$, $1.409r_{0}$ respectively. Note that for $r_c=1.235r_0$, $C_3(T)$ (the blue curve) dominates very distinctively over all other species at low temperatures indicating a three-fold symmetric orientational correlation of a perfect HS in particles positions. However, for $r_c = 1.409r_{0}$ the three-fold symmetry is observed in case of (a) $\rho=0.747$, (b) $\rho=0.792$ while the four-fold symmetry of a RS is observed for (c) $\rho=0.804$. The strong orientational correlation present in the system leads to a strong glass.

\subsection{Liquid-Solid Transition}
\begin{figure}[!h]
\begin{center}
\hskip -0.5cm
\includegraphics[width=3.4 in]{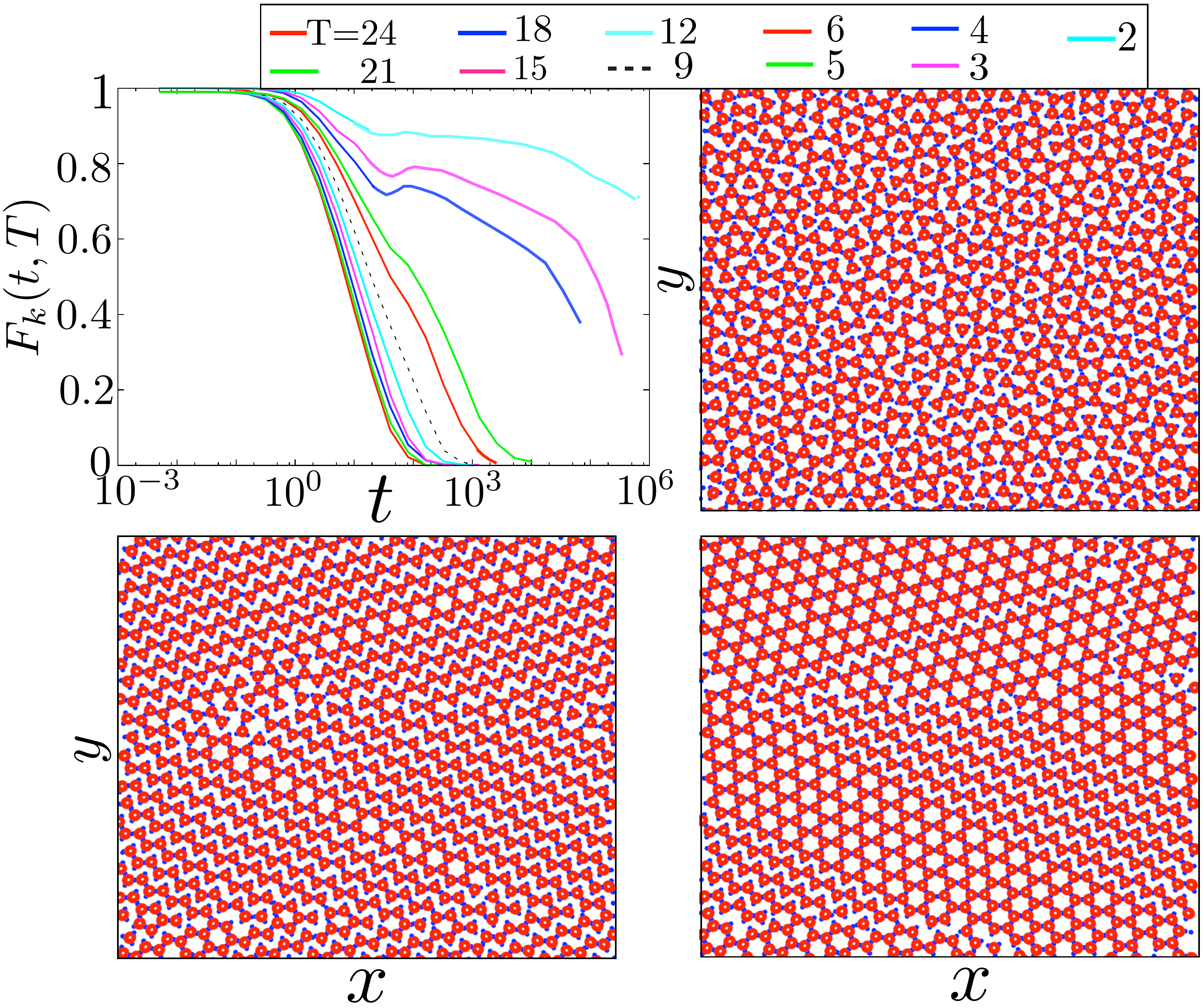}
\caption{(a)Self part of the intermediate scattering function $F_{k}(t,T)$ as a function of time $t$ for different temperatures $T$ at $\rho=0.758$. Snapshot picture for (b) $T=5.0$, (c) $T=4.0$, and (d) $T=2.0$. The system undergoes a liquid-solid transition near $T=4.0$. Note that the $\beta$ -decay plateau is missing in the correlation functions.}
\label{Fig.76}
\end{center}
\end{figure}
Glass transition in a liquid is a dynamical property as it strongly depends on how fast the material is cooled. If the system is quenched very slowly, it may solidify to it's ground state. We observe this phenomena in our system. We quenched the system from $T=24.0$ extremely slowly to its HS ground state structure at $\rho=0.758$. In this protocol we ran the simulations upto a maximum of $10^9$ MD steps with the same step size as before. The resulting correlation functions are shown in Fig.\ref{Fig.76}a and the microscopic configurations at (b) $T=5.0$, (c) $T=4.0$, and (d) $T=2.0$ are also shown. The signature of a liquid-solid transition is very evident from these figures. $\beta$ -decay plateau which is a glass-forming signature, is missing in the correlation functions at low temperatures.

\section{Conclusions and future directions}
\label{summary}
Polymorphism, which arises mainly from 
the complicated patchy nature of the interaction potential, is common in network 
forming liquids like water and SiO$_2$; water having eleven distinct solid phases!
It is well known that water shows many interesting anomalous dynamical behaviours 
due to these large number of low temperature solid phases including glassy states, 
liquid-liquid phase transition etc. In this model we show how the existence of
multiple low temperature solid states gives rise to very interesting dynamical 
behaviour including glass like dynamics. According to different theories of glass
transition like Random First Order Transition (RFOT) theory \cite{rfot1,rfot2} and 
frustration limited domain \cite{kivelson,tarjus1,chayes,nussinov,tarjus2}, 
it is often assumed that slowing down in the dynamics is caused by existence of 
multiple minima in the energy landscape at lower temperatures. In RFOT it is assumed 
that there are exponentially large number of metastable states below a dynamical transition 
temperature and system gets trapped at one such minima at still lower temperature.
On the other hand in the frustration limited domain theory it is again assumed 
that system wants to form locally preferred ordered domains but often that 
structure can not fill the space efficiently leading to frustration in the 
whole system.        

In this paper we construct a model for patchy colloids with a very rich phase 
behaviour consisting of three different crystalline ground states as well as 
isotropic and networked liquid phases for different choices of parameters. We observe that due to geometrical frustration the system cannot equilibrate to one of its homogeneous low temperature equilibrium states but gets arrested in a polycrystalline or amorphous structures with a prominent $\beta$ relaxation plateau in the intermediate scattering function -- the hallmark of a glass transition. To obtain such an outcome, the high temperature liquid needs to be cooled at sufficiently fast cooling rate and parameters need to be chosen such that more than one crystalline state is equally stable.  The existence of multiple low temperature crystalline states therefore directly leads to glassy dynamics and amorphous structures. In this case, the system, like other network forming liquids, forms a strong glass with Arrhenius relaxation due to the presence of strong orientational correlations. On the other hand, when cooled sufficiently slowly for a choice of parameters such that a well defined, unique, crystalline ground state exists we obtain a liquid to crystal transition with a vanishing $\beta$ relaxation regime.   

Our work suggests that most glass forming liquids may have multiple competing ground states with large and complex unit cells -- a fact which needs to be revealed by detailed studies in the future. In this context a generalisation of our model to study a patchy colloids with randomly distributed patches will be interesting. This random patchy colloidal model might have many crystalline ground states with large unit cells and might help us unearth some of the puzzles of the glass transition. Variation of the nature of the glass transition, strong or fragile, with the number and strength of the patches is also of interest.

In future it will also be very interesting to study the stability of these glassy 
states and see how they respond against an external deformation. While non-
metallic glasses like silicate glasses\cite{silicate-glass} shows brittle 
nature, metallic glasses\cite{Ductile-BMG} show ductile failure. To this 
end a rheological study of the low temperature quenched glassy states as 
well as crystalline honeycomb solid and triangular solid state is required to 
be investigated. The question of how the initial configurations being either 
crystalline, polycrystalline or amorphous influence the flow behaviour of the 
system and whether the flow behaviour is identical to that of glass will be very 
useful to understand many interesting rheological properties of the network
forming glassy liquids. 

Calculations along both these directions are in progress and will be published elsewhere.

\acknowledgements
C.M. acknowledges support from a CSIR SRF and from TCIS. Numerous discussions 
over the years with Biman Bagchi on this and many other subjects are gratefully acknowledged.

\section{Appendix}
The interaction between two $type-1$ particles $(U_{1,1})$ is a Hamaker interaction\cite{Hamaker} between two large sized colloidal particles given by:
\begin{eqnarray}
U_{A}(r) = \frac{A_{1,1}}{6}\Big[\frac{2a_{1}a_{2}}{r^{2}-(a_{1}+a_{2})^{2}} + \frac{2a_{1}a_{2}}{r^{2}-(a_{1}-a_{2})^{2}} \,\,\,\,\,\,\,\,\,\,\,\,\,\,\,\,\,\,\,\,\,\,\nonumber \\
+ ln\left( \frac{r^{2}-(a_{1}+a_{2})^{2}}{r^{2}-(a_{1}-a_{2})^{2}} \right) \Big]\,\,\,\,\,\,\,\,\,\,\,\,\,\,\,\,\,\,\,\,\,\,
\end{eqnarray}
\begin{eqnarray}
U_R(r) = \frac{A_{1,1}}{37800}\frac{\sigma_{1,1}^6}{r}\Big[\frac{r^2 - 7r(a_1 + a_2) + 6(a_1^2 + 7a_1a_2 + a_2^2)}{(r - a_1 - a_2)^7} \nonumber \\
+ \frac{r^2 + 7r(a_1 + a_2) + 6(a_1^2 + 7a_1a_2 + a_2^2)}{(r + a_1 + a_2)^7} \nonumber \\
- \frac{r^2 + 7r(a_1 - a_2) + 6(a_1^2 - 7a_1a_2 + a_2^2)}{(r + a_1 - a_2)^7} \nonumber \\
- \frac{r^2 - 7r(a_1 - a_2) + 6(a_1^2 - 7a_1a_2 + a_2^2)}{(r - a_1 + a_2)^7}\Big]\nonumber \\
U_{1,1}(r) = U_A + U_R , \,\,\,\,\,\,\,\,\,\,\,\,\,\,\,\,\,\,\,\,\,\,\,\,\,\, r<r_{c1,1} \,\,\,\,\,\,\,\,\,\,\,\,\,\,\,\,\,\,\,\,\,\,\,\,\,\, \nonumber \\
U_{1,1}(r) = 0 , \,\,\,\,\,\,\,\,\,\,\,\,\,\,\,\,\,\,\,\,\,\,\,\,\,\,\,\,\,\,\,\,\,\,\,\,\,\,\,\,\,\,\,\,\ r\ge r_{c1,1} \,\,\,\,\,\,\,\,\,\,\,\,\,\,\,\,\,\,\,\,\,\,\,\,\,\, 
\label{eqn:u11}
\end{eqnarray}
where $A_{11}$ is the Hamaker constant, $a_{1}$ and $a_{2}$ are the radii of the two colloidal particles, and and $r_{c}$ is the cutoff. In our case $a_{1} = a_{2} = a$. This equation is derived in\cite{Hamaker}. It results from describing each colloidal particle as an integrated collection of Lennard-Jones particles of size sigma. 
The interaction between a $type-1$ particle and a $type-2/type-3$ particle is the interaction between a large size colloid particle and a solvent particle and is given by:
\begin{eqnarray}
U_{1,2/3}(r) &=& \frac{2a^3\sigma_{1,2/3}^3A_{1,2/3}}{9(a^2 - r^2)^3}\Big[1-
\,\,\,\,\,\,\,\,\,\,\,\,\,\,\,\,\,\,\,\,\,\,\,\,\,\,\,\,\,\,\,\,\,\,\,\,\,\,\,\,\,\,\,\,\,\,\,\,\,\,\,\, \nonumber \\ 
&&\frac{(5a^6+45a^4r^2+63a^2r^4+15r^6)\sigma_{1,2/3}^6}{15(a-
r)^6(a+r)^6}\Big], \nonumber \\
& &\,\,\,\,\,\,\,\,\,\,\,\,\,\,\,\,\,\,\,\,\,\,\,\,\,\,\,\,\,\,\,\,\,\,\,\,\,\,\,\,\,\,\,\,\,\,\,\,\,\,\,\,\,\,\,\,\,\,\,\,\,\,\,\,\,\,\,\,\,\,\,\,\,\,\,\,\,\,\,\ r<r_{c1,2/3} \nonumber \\
&=& 0 ,\,\,\,\,\,\,\,\,\,\,\,\,\,\,\,\,\,\,\,\,\,\,\,\,\,\,\,\,\,\,\,\,\,\,\,\,\,\,\,\,\,\,\,\,\,\,\,\,\,\,\,\,\,\,\,\,\,\,\,\,\,\,\,\,\,\,\,\,\,\,\,\,\,\,\, r\ge r_{c1,2/3} \nonumber \\
\label{eqn:u123}
\end{eqnarray}
where $A_{1,2/3}$ are the Hamaker constant, $a$ is the radius of the colloidal particle. This formula is derived from the colloid-colloid interaction, letting one of the particle sizes go to zero.
The interaction between two $type-2$ particles is given below:
\begin{eqnarray}
U_{2,2}(r) &=& \frac{A_{2,2}}{36}\Big[(\frac{\sigma_{2,2}}{r})^{12} - (\frac{\sigma_{2,2}}{r})^6\Big], r<r_{c2,2} \nonumber \\
&=& 0 , \,\,\,\,\,\,\,\,\,\,\,\,\,\,\,\,\,\,\,\,\,\,\,\,\,\,\,\,\,\,\,\,\,\,\,\,\,\,\,\,\,\,\,\,\,\,\,\,\,\,\,\,\,\,\,\,\,\,\, r\ge r_{c2,2}
\label{eqn:u22}
\end{eqnarray}
and two $type-3$ particles or a $type-2$ particle and a $type-3$ particle interact with simple simple LJ interaction:
\begin{eqnarray}
U_{m,n}(r) &=& 4A_{m,n}\Big[(\frac{\sigma_{m,n}}{r})^{12} - (\frac{\sigma_{m,n}}{r})^6\Big], r<r_{cm,n} \nonumber \\
&=& 0 , \,\,\,\,\,\,\,\,\,\,\,\,\,\,\,\,\,\,\,\,\,\,\,\,\,\,\,\,\,\,\,\,\,\,\,\,\,\,\,\,\,\,\,\,\,\,\,\,\,\,\,\,\,\,\,\,\,\,\,\,\,\,\, r\ge r_{cm,n} 
\label{eqn:u33}
\end{eqnarray}
Where, $m, n = 2, 3$ except $m=n=2$.
 
The sizes and interaction strengths between $type-1$, $type-2$ and $type-3$ particles are tabulated below. We choose the size of the $type-1$ particle $a=7.0\sigma_{11}$.
\begin{table}[h!]
\caption{Interaction parameters}
\begin{center}
\begin{tabular}{c c c}
\hline\hline \\ [1ex]
$A_{1,1} = 1.0$ &  $A_{2,2} = 1140.0$ & $A_{3,3} = 0.2$\\  [1ex]
$A_{1,2} = 20.0$ & $A_{1,3} = 1.0$ & $A_{2,3} = 1.0$\\  [1ex]
$\sigma_{1,1} = 1.0$ & $\sigma_{2,2} = 0.7989$ & $\sigma_{3,3} = 7.029$\\  [1ex]
$\sigma_{1,2} = 1.0023$ & $\sigma_{1,3} = 1.05$ & $\sigma_{2,3} = 6.6734$\\  [1ex]
$r_{c1,1} = 21.0$ & $r_{c2,2} = 4.0$ & $r_{c3,3} = 14.0$\\  [1ex]
$r_{c1,2} = 12.0$ & $r_{c1,3} = 16.0$ & $r_{c2,3} = 14.0$\\  [4ex]
\hline
\end{tabular}
\end{center}
\label{table:IP}
\end{table}

\end{document}